\documentclass[journal]{IEEEtran}
\usepackage{graphicx}
\usepackage{dcolumn}
\usepackage{bm}
\usepackage{caption}
\usepackage{comment}
\usepackage{subcaption}
\usepackage{xcolor}
\usepackage{physics}
\usepackage{ulem}
\usepackage{placeins}
\usepackage{circuitikz}
\usepackage{lineno}
\usepackage{mathrsfs}
\usepackage{citesort}

\newcommand\BibTeX{{\rmfamily B\kern-.05em \textsc{i\kern-.025em b}\kern-.08em
T\kern-.1667em\lower.7ex\hbox{E}\kern-.125emX}}

\begin{document}

\hyphenation{op-tical net-works semi-conduc-tor}
 \newcommand{\shankernote}[1]{\textcolor{red}{[SB: #1]}}
 \newcommand{\zanenote}[1]{\textcolor{purple}{[ZC: #1]}}
 \newcommand{\omkarnote}[1]{\textcolor{green}{[OHR: #1]}}

\title{Port Parameter Extraction Based Self Consistent Coupled EM-Circuit FEM Solvers}
\author{O. H. Ramachandran,~\IEEEmembership{Student Member,~IEEE,}
        S. O'Connor,~\IEEEmembership{Student Member,~IEEE,}
        Z. D. Crawford,~\IEEEmembership{Student Member,~IEEE},
        L. C. Kempel,~\IEEEmembership{Fellow,~IEEE},
        B. Shanker,~\IEEEmembership{Fellow,~IEEE}
        }

\maketitle

\begin{abstract}
Self consistent solution to electromagnetic (EM)-circuit systems is of significant interest for a number of  applications. This has resulted in exhaustive research on means to couple them. In time domain, this typically involves a tight integration (or coupling) with field and non-linear circuit solvers. This is in stark contrast to coupled analysis of linear/weakly non-linear circuits and EM systems in frequency domain. Here, one typically extracts equivalent port parameters that are then fed into the circuit solver. Such an approach has several advantages; (a) the number of ports is typically smaller than the number of degrees of freedom, resulting in cost savings; (b) is circuit agnostic; (c) can be integrated with  a variety of device models. Port extraction is tantamount to obtaining impulse response of the linear EM system. In time domain, the deconvolution required to effect this is unstable. Recently, a novel approach was developed for time domain integral equations to overcome this bottleneck. We extend this approach to time domain finite element method, and demonstrate its utility via a number of examples; significantly, we demonstrate that self consistent solutions obtained using either a fully coupled or port extraction is identical to the desired precision for non-linear circuit systems. This is shown within a nodal network. We also demonstrate integration of port extracted data directly with drift diffusion equation to model device physics.

\end{abstract}

\maketitle


\section{Introduction}\label{sec1}

The combined simulation of full-wave electromagnetics solvers with circuit subsystems
are of considerable interest in a number of applications, including analysis of 
shielded or packaged systems,
active antenna design, small signal analysis of active devices and 
design of high speed interconnects \cite{kong_FDTD,li_book,interconnect_1}.
It is driven by advances in design techniques that permit fabrication of complex RF devices with active elements,
making it important to characterize  radiative  coupling effects 
early in the design process.
To that end, extensive work has been done in developing coupled Maxwell-circuit solvers in time and frequency domain, with the current state of the art utilizing finite element \cite{wang_jin_2008} or integral equation methods \cite{ie_cite2,ie_cite1}. Here, we restrict our discussion to transient analysis as they better resolve strongly nonlinear systems.

The predominant approach to transient analysis of EM-circuit system is to solve the system self consistently \cite{jin_riley}. This involves solving both the linear and non-linear system at every time step. Obviously, the tight integration implies that the solution is not circuit agnostic. Alternatives that have been explored is to use frequency domain methods to construct a transient ``impulse response'' that take the form either as $RC$ extraction \cite{Kao_2001_parasitic}) or $S$ parameter methods \cite{Antonini_S_param_2005}. The extracted response is readily incorporated into a circuit simulator. While this approach is somewhat effective, the advantages and limitation are apparent; (a) the approach is independent of the circuit system; (b) given the bandwidth of excitation, the harmonics generated due to non-linearity and need to capture early time behavior, the number of frequency samples necessary can be very high; and (c) often only a subset of these frequencies are used. In weakly non-linear systems or when the coupling is not strong the errors accrued may be tolerable. When analysis of the circuit system is possible using frequency domain techniques (harmonic balance) under the assumption of weakly non-linear systems, one often takes recourse to using a Schur complement approach to couple EM to circuit systems \cite{shi_hb,scott_pe}. In addition to being circuit agnostic, this is computationally more efficient as there are fewer ports than number of spatial degrees of freedom. 

It follows, that developing such a methodology for transient analysis will have the same benefits, in addition to potential integration with multiphysics codes that model device physics. Extracting port parameters of the EM system is analogous  to computing its numerical impulse response. Doing this in time domain is  challenging because of known  instabilities associated with deconvolution \cite{charland_98}. A recently proposed technique \cite{scott_pe} for solving coupled circuits with time 
domain integral equation (TDIE) solvers overcame this fundamental bottleneck. Extending this technique to finite element based solvers involves several changes in the extraction process.
First, the extracted signal manifests itself as a transient admittance in the circuit system as opposed to an impedance in \cite{scott_pe}. As a result, the feed model used is changed to a current probe as opposed to a delta-gap feed, leading to differences both in coupling and in the extraction process.
Finally, using a finite element scheme allows for integration with different set differential equations  used to model the device subsystem; in this paper, we demonstrate this capability via coupling with a non-linear Drift-Diffusion equations to model a Schottky diode. The specific details involved will be covered in depth in Section \ref{sec:extraction}.

The principal contributions of this paper are (a) the development of a method for extracting transient port parameters in the EM-circuit interface for finite element systems, (b) the demonstration that solutions obtained through this method are identical to those obtained using a fully coupled solution to solver precision, and (c) integration with device specific differential equations. Furthermore, we also briefly demonstrate the implementation of a Perfectly Matched Layer (PML) system for mixed finite element electromagnetic solvers. Via numerous examples, we will demonstrate the application of these for analysis of linear and nonlinear circuits coupled to EM systems. 
Where possible, we will show comparison against data that exists in the literature (either measured or modeled). 
We note that while our results are obtained using a implicit mixed FEM system, the prescribed procedure is applicable to the traditional wave equation solvers. Our rationale for using mixed FEM as opposed to the usual wave equation is that the latter has a time growing null-space of the form $t\grad \psi (\vb{r})$, whereas the former has a null space of the form $\grad \psi (\vb{r})$. In mixed FEM, the magnitude of the null-space excited depends on the threshold used for the iterative solver. That said, it should noted that a gauging constraint as described in \cite{wang2010application,venkatarayalu2006removal,li2016finite,o2021quasi} eliminates this null space. 
Finally, the nonlinearities are assumed to be lumped or pointwise. While the proposed method can potentially be used to isolate small regions of continuous nonlinear materials, we relegate this to a future paper.

The rest of the paper will be structured as follows:
Section \ref{sec:formulation} will detail the implementation of the mixed finite 
element system for the EM system and the MNA solver for the attached circuits;
Section \ref{sec:extraction} will describe the technique involved in extracting
a set of transient port parameters from the EM system and using it to solve the 
coupled problem; Finally,  Section \ref{sec:results} will contain a set of numerical
examples to both validate the method and demonstrate its efficacy.

\section{Formulation}\label{sec:formulation}
\subsection{Problem Statement}

Consider an object $\Omega_\text{EM}\in\mathbf{R}^{3}$ bounded by a surface $\partial\Omega_\text{EM}$, 
that describes the geometry of an electromagnetic object containing
$N_{p}$ ports, each associated with a lumped circuit subsystem.
The currents flowing across these ports are collectively represented as $\mathbf{J}^{\text{CKT}}(\mathbf{r},t)$ with $\mathbf{r}\in\Omega_{EM}$.
We assume that any voltage sources in the circuit system are bandlimited
to some frequency range $\left[f_{\text{min}},f_{\textrm{max}}\right]$ with $f_{\text{min}}>0$.
Furthermore, we assume that the amplitude of these sources are zero when $t\leq0$.
\subsection{Modelling Framework}
We construct a Maxwell solver following a mixed finite element scheme using Whitney edge and face basis functions 
$\mathbf{E}(\mathbf{r},t) = \sum_{i=1}^{N_{e}}e_{i}(t)\mathbf{W}_{i}^{1}(\mathbf{r})$
and $\mathbf{B}(\mathbf{r},t) = \sum_{i=1}^{N_{f}}b_{i}(t)\mathbf{W}_{i}^{2}(\mathbf{r})$ ,
where $N_{e}$ and $N_{f}$ are the number of edges and faces respectively of the tetrahedral mesh to discretize the domain; see \cite{crawford} and references therein.
The EM unknowns are represented in time as $\mathbf{e}(t)=\sum_{j=1}^{N_{t}}e_{j}N(t-t_{j})$ and $\mathbf{b}(t)=\sum_{j=1}^{N_{t}}b_{j}N(t-t_{j})$, and tested by $W(t-t_{i})$. Both of these functions are defined in \cite{newmark}.
A Newmark-$\beta$ time stepping stencil with $\gamma = 0.5$ 
and $\beta = 0.25$ is used to solve for $\mathbf{e}(t)$ and $\mathbf{b}(t)$ and an appropriately configured PML to truncate the computational domain.
Contemporary implementations of PML systems follow the general framework first outlined by Berenger \cite{BERENGER1994185} with more recent additions, including the use of stretched coordinates \cite{PML}. The implementation of these systems is done by either directly evaluating the convolutions resulting from the use of a stretched coordinate system or defining and solving for two auxiliary variables in addition to the regular field unknowns to achieve the same effect.
The PML implementation used in this paper directly evaluates the convolution integrals. To do so, we define a stretched coordinate system via the following transform
\begin{equation}
    \mathbf{\Lambda}(\omega) =
    \left(
    \begin{matrix}
    \frac{s_{y}s_{z}}{s_{x}} & 0 & 0 \\
    0 & \frac{s_{x}s_{z}}{s_{y}} & 0 \\
    0 & 0 & \frac{s_{x}s_{y}}{s_{z}}
    \end{matrix}
    \right)
\end{equation}
with $s_{i} = 1+\frac{\sigma_{i}}{j\omega\epsilon_{0}}$ to match the absorbing layers to free space. 
Here, $\sigma_{i}$ are the components of an anisotropic conductivity that governs the field loss. Stretching coordinates in this manner alters Maxwell's equations as follows in frequency domain:
\begin{equation}\label{eq:PML_freq}
    \begin{split}
        \mathbf{\Lambda}(\omega)^{-1}\cdot\curl \mathbf{E}(\mathbf{r},\omega) &= -\mathbf{\Lambda}(\omega)^{-1}\cdot j\omega\mathbf{B}(\mathbf{r},\omega) \\
        \curl \mu^{-1}\mathbf{\Lambda}(\omega)^{-1}\cdot\mathbf{B}(\mathbf{r},\omega) &= \mathbf{J}(\mathbf{r},\omega) + \epsilon(\mathbf{r})\mathbf{\Lambda}(\omega)\cdot j\omega\mathbf{E}(\mathbf{r},\omega)
    \end{split}
\end{equation}
Obtaining a time marching scheme involves inverse Fourier transforming \eqref{eq:PML_freq} to obtain
\begin{subequations}\label{eq:PML_TD}
    \begin{align}
        \mathbf{L}_{2}(t)*\curl\mathbf{E}(\mathbf{r},t) &= -\mathbf{L}_{2}(t)*\frac{\partial \mathbf{B}(\mathbf{r},t)}{\partial t} \label{eq:PML_1} \\
        \curl \mu^{-1}\mathbf{L}_{2}(t)*\mathbf{B}(\mathbf{r},t) &= \mathbf{J}(\mathbf{r},t) + \epsilon_{0}\mathbf{L}_{1}(t)*\mathbf{E}(\mathbf{r},t) \label{eq:PML_2}
    \end{align}
\end{subequations}
where $L_{1}(t) = \mathscr{F}^{-1}\left(j\omega\mathbf{\Lambda(\omega)}\right)$ and $L_{2} = \mathscr{F}^{-1}\left(\mathbf{\Lambda(\omega)}^{-1}\right)$.
We discretize these equations by testing \eqref{eq:PML_1} with a $\mathbf{W}^{2}(\mathbf{r})$ basis function and \eqref{eq:PML_2} with $\mathbf{W}^{1}(\mathbf{r})$.
Furthermore, the convolution terms are evaluated as done in \cite{PML}.

The behavior of the attached devices at each port can be described generally by operators $\mathcal{D}$, $\mathcal{F}$ and couples to the EM system through $\mathcal{C_{\text{CKT}}}$, forming
\begin{equation}\label{eq:device_op}
    \mathcal{D}\circ\left[\mathbf{J}^{\text{CKT}}(\mathbf{r},t),\mathbf{e}(t)\right] +\mathcal{C_{\text{CKT}}}\circ\left[\mathbf{e}(t)\right]= \mathcal{F}\circ\left[\mathbf{e}(t)\right].
\end{equation}
Where $\mathcal{D}$ and $\mathcal{F}$ are general nonlinear operators and $\mathcal{C}_{\text{CKT}}$ is a coupling operator that relates quantities in the EM system to those in the attached device. For the results presented in this work, we restrict $\mathcal{D}$ to either be an circuit network implemented through Modified Nodal Analysis \cite{mna}; used entirely by itself or in conjunction system governed by a set of Drift-Diffusion equations to model diodes \cite{scharfettergummel1969,kurataDdiffusion1972}.
In the case of the former, we temporally represent the voltage and circuit unknowns using a $\mathit{p}$th order
backward Lagrange interpolation function $L_{p}(t-t_{i})$. We note that this choice of representation is what is commonly used in contemporary implementations of MNA, but are in no way the only feasible choice. Upon using our chosen representation, we obtain the following system
\begin{equation}\label{eq:ckt_time_marching}
	\begin{split}
		\mathbf{Y}\mathbf{V}^{\text{CKT}}(t) = 
		\mathbf{f}^{\text{CKT}}(t) + \mathbf{f}_{nl}^{\text{CKT}}(\mathbf{V}^{\text{CKT}},t).
	\end{split}	
\end{equation}
which is subsequently delta tested to obtain a time marching scheme.
Here $\mathbf{V}^{\text{CKT}}$ is a vector containing both the nodal voltages and
branch currents in the circuit,
$\mathbf{f}^{\text{CKT}}$ and $\mathbf{f}^{\text{CKT}}_{nl}$ refer to the 
excitations due to linear and nonlinear components.
The linearized form in \eqref{eq:ckt_time_marching} can be solved at each timestep 
using a multi-dimensional Newton-Raphson scheme similar to \cite{wang_jin_2008}.
Similarly, when employing a drift diffusion operator to model diodes in the system, the currents due to electrons and holes $\mathbf{J}_{n}(\mathbf{r},t)$ and $\mathbf{J}_{p}(\mathbf{r},t)$ running through the device are related to carrier densities $n(\mathbf{r},t)$, $p(\mathbf{r},t)$ and potential $\phi(\mathbf{r},t)$ through
\begin{subequations} \label{eq:DDESchottky}
\begin{equation}
\begin{split}
    \mathbf{J}_{n}(\mathbf{r},t) = q D_{n}\grad n(\mathbf{r},t) +q\mu_{n}(\mathbf{E}(\mathbf{r},t))n(\mathbf{r},t)\nabla \phi(\mathbf{r},t)
\end{split}
\end{equation}
\begin{equation}
\begin{split}
    \mathbf{J}_{p}(\mathbf{r},t) = - q D_{p}\grad p(\mathbf{r},t) +q\mu_{p}(\mathbf{E}(\mathbf{r},t))p(\mathbf{r},t)\nabla \phi(\mathbf{r},t)
\end{split}
\end{equation}
where $\mu_{n}$ and $\mu_{p}$ are field dependent mobility rates for the electrons and holes respectively; and $D_{p}$ and $D_{n}$ are corresponding diffusion coefficients. The currents and carrier densities are further related through a set of continuity equations:
\begin{equation}
\begin{split}
    \frac{\partial n(\mathbf{r},t)}{\partial t} = \frac{\nabla \mathbf{J}_{n}(\mathbf{r},t)}{q} - R + G 
\end{split}
\end{equation}
\begin{equation}
\begin{split}
    \frac{\partial p(\mathbf{r},t)}{\partial t} = -\frac{\nabla \mathbf{J}_{p}(\mathbf{r},t)}{q} - R + G
\end{split}
\end{equation}
where $R$ and $G$ respectively denote the electron-hole recombination and the collision ionization rates.
Finally, the carrier densities are related to the potential through Poisson's equation
\begin{equation}
    \nabla\cdot(\epsilon \nabla\phi (\mathbf{r},t)) = -q\left(p(\mathbf{r},t)-n(\mathbf{r},t)+N_{t}(\mathbf{r},t)\right)
\end{equation}
\end{subequations}
where $N_{t}(\mathbf{r},t)$ refers to the doping concentration. Solution to the drift diffusion system can be obtained by discretizing \eqref{eq:DDESchottky} using an appropriate finite element or finite difference method; see \cite{scharfettergummel1969,kurataDdiffusion1972,junquan2012Ddiffusion} and the references therein for a detailed analysis. The results presented in this paper only involve 1D drift-diffusion systems and as a result we discretize \eqref{eq:DDESchottky} using a corresponding 1D finite element system.

We describe the interaction between EM and device subsystems in two parts. First, we consider a device system modelled using MNA. In this instance, the quantities involved in the device system are voltages and currents, which need to be related to fields and current densities in the EM system. Specifically, if the $k$th FEM edge (denoted by $\mathbf{l}_{k}$) is attached to the $j$th circuit subsystem, the current impressed on the EM system is given by
\begin{equation}\label{eq:circ_to_em}
	\begin{split}
	\langle W(t-t_{i}),J^{\text{CKT}}_{k}(t)\rangle &= \langle W(t-t_{i}),I^{\text{CP}}_{j}(t)\int_{|\mathbf{l}_{k}|}
	\hat{\mathbf{l}}_{k}\cdot \mathbf{W}^{1}_{k} d\mathbf{r}\rangle \\
	&= \langle W(t-t_{i}),I^{\text{CP}}_{j}(t) C_{kj}\rangle    
	\end{split}
\end{equation}
with $C_{kj}$ denoting a coupling coefficient that relates quantities in the device subsystem to the EM solver. Furthermore, $I^{\text{CP}}_{j}(t)$ refers to the magnitude of the current impressed by the circuit subsystem over the coupling edge $\mathbf{l}_{k}$.
Similarly, the voltage across the coupling branch can be related to the electric field across the $k$th FEM edge
\begin{equation}\label{eq:em_to_circ}
	\begin{split}
		\langle \delta(t-t_{i}),V^{\text{CKT}}_{j}(t)\rangle &= \langle \delta(t-t_{i}),e_{k}(t) \int_{|\mathbf{l}_{k}|} \hat{\mathbf{l}}_{k} \cdot \mathbf{W}^{1}_{k}(\mathbf{r}) d\mathbf{r}\rangle \\
			   &= \langle \delta(t-t_{i}),e_{k}(t) C_{jk}\rangle.
	\end{split}
\end{equation}
$C_{jk}$ here likewise denotes a coupling coefficient that relates quantities in EM solver to the device. We observe from \eqref{eq:em_to_circ} and \eqref{eq:circ_to_em} that our choice of testing/representation functions leads the two coupling coefficients to be identical. For a drift diffusion setup, the electric field at the location of the port is related to the electron and hole mobilities $\mu_{n}(\mathbf{E}(\mathbf{r},t))$ and $\mu_{n}(\mathbf{E}(\mathbf{r},t))$ respectively. Since the devices are assumed to be lumped, $\mathbf{E}(\mathbf{r},t)$ at the location of the port can be used directly to compute the carrier mobilities, since the field is assumed to be spatially constant within the device. Likewise, we can use $\mathbf{J}_{p}(\mathbf{r},t)$ and $\mathbf{J}_{n}(\mathbf{r},t)$ are to construct the net current passing through the diode, which can then be reintroduced to the EM system following \eqref{eq:circ_to_em}.

\section{Extraction of the Numerical Impulse Response}\label{sec:extraction}
The computational bottlenecks involved with solving a coupled system as described in the previous Section are twofold:
(1) Resolving nonlinear elements in the circuit system involves performing a solve of the combined matrix equation
and (2) changing any of the attached circuit subsystems would require the coupled problem to be solved again, despite the EM system remaining unaltered.
A potential way to exploit the linearity of the EM system is to extract its impulse
response at each EM-circuit interface and use it through \eqref{eq:em_to_circ} in the circuit solve.
Unfortunately, it is well known that deconvolution required to implement this is 
unstable \cite{charland_98}.

The key insight in the method proposed herein is as follows. 
The current deposited on a given port edge is represented in time through a linear combination of $N_{t}$ basis functions.
As a result, given the EM response due to a single temporal basis 
function, we can exploit the linearity of Maxwell's equations and reconstruct the 
field anywhere in the system.
Since this sequence of operations only involves reconstructing the current at a given
port in terms of basis functions by which it is represented in the coupled solve, 
the respective fields computed by both methods should be numerically 
indistinguishable.

With $p(q)$ denoting the set of FEM edges associated with the port $q$, we define an excitation vector $\mathbf{e}_{q}(t)$ defined as follows
\begin{equation}
    e_{q,k}(t) =
    \begin{cases}
        N(t-t_{\delta}) & k\in p(q) \\
        0 & \text{otherwise}
    \end{cases}
\end{equation}
where $\delta$ is the timestep at which the excitation is applied.
$\mathbf{e}_{q}$ is then used to define the forcing function 
$\mathbf{J}^{\text{CKT}}(t)$ through \eqref{eq:circ_to_em} with $J^{\text{CKT}}(t) = C_{kq}\mathbf{e}_{q}(t)$. This function is then applied to the RHS of \eqref{eq:PML_TD} to obtain a solution vector $\mathbf{x}^{q}$.
In order to solve the device equations, however, we only require the coefficients
associated with each port, allowing us to construct a matrix $G_{kq} = x^{k,p(q)}$ of dimensions $N_{p}\times N_{p}\times N_{t}$.
Each column of $G$ represents a discrete impulse response for a pulse centered at the edge $p(q)$ measured at the $k$th edge.
As a result, constructing the electric field at port $k$, in response to an arbitrary set of currents can be done by simply summing the convolutions of $G_{kq}$ with $I_{q}$ for each attached circuit port. The reconstructed fields can then be related using the appropriate coupling equations to quantities the device subsystems.
For instance, if the attached port is governed through MNA, the voltage across the $j$th port
$V^{\text{CKT}}_{j} (t)$ in \eqref{eq:em_to_circ} can now be written in terms of $I^{\text{CP}} (t)$
\begin{equation}\label{eq:Vjk_from_G}
\begin{split}
		V^{\text{CKT}}_{j}(t_{i})
			   &= \langle \delta(t-t_{i}),C_{jk}\sum_{q=1}^{N_{p}}G_{kq}(t)*I_{q}^{\text{CP}}(t) \rangle.
	\end{split}
\end{equation}
yielding a standalone matrix equation for the device system.
\section{Results}\label{sec:results}
The numerical experiments presented in this section will be organized as follows: 
Sections \ref{sec:linear_benchmark} and \ref{sec:nonlinear_benchmark} will 
compare results obtained using the port extraction technique described in 
Section \ref{sec:extraction} against existing results in the literature for both linear and nonlinear circuit systems. 
Section \ref{sec:showing_off} will highlight three key facts about the proposed method; first, we demonstrate that the solutions obtained through port extraction are numerically identical to their fully coupled counterparts; second, we show that the extraction procedure is circuit agnostic; and finally, we compare the complexity of the port extracted solve to a traditional fully coupled setup.

For the results presented in the remainder of this section, $N_{t}$ denotes the number of timesteps that the simulation is run over and $N_{\text{EM}}$, $N_{\text{CKT}}$ denote the numbers of EM and circuit unknowns respectively in the system.
Unless specified otherwise, voltage sources are defined using  $v(t) = \cos(2\pi f_{0} t)e^{-t^{2}/2\sigma^{2}}$ where $\sigma = 3\times\left(2\pi f_{\text{bw}}\right)^{-1}$, with $f_{\text{max}}=f_{0}+f_{\text{bw}}$.
The timestep size $\Delta_{t}=\left(30f_{max}\right)^{-1}$. 
Finally, GMRES was used to solve the system iteratively to a tolerance of $10^{-12}$.

\subsection{Input Impedance of a Monopole Antenna}
\begin{figure}
\begin{subfigure}{0.5\textwidth}
    \centering
    \includegraphics[width=0.8\linewidth]{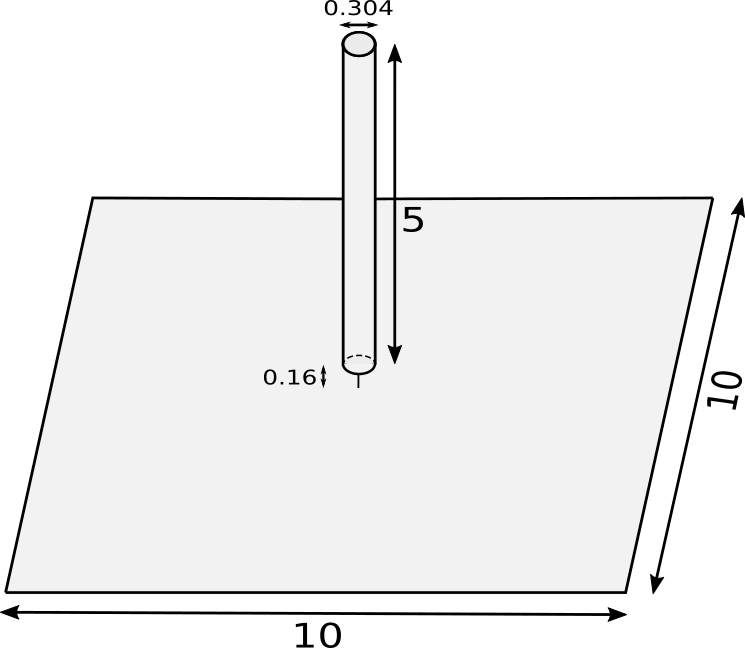}
    \caption{Geometry of the monopole system. The dimensions are in cm.}
    \label{fig:cyl_monopole}
\end{subfigure}
\begin{subfigure}{0.5\textwidth}
    \centering
    \includegraphics[width=\linewidth]{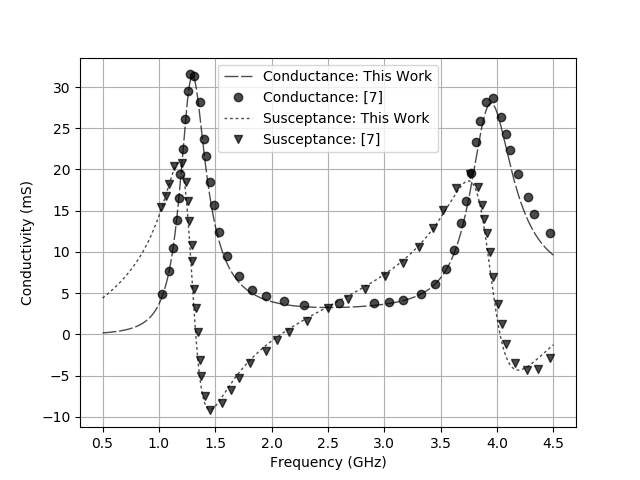}
    \caption{Plots of the input admittance in S measured from 0.5 GHz to 4.5 GHz compared to existing results in the literature \cite{jin_riley}.}
    \label{fig:jin_monopole}
\end{subfigure}
\caption{Description of the geometry and obtained results for the cylindrical monopole antenna.}
\end{figure}
In this first example, we validate our technique by analysing a cylindrical monopole suspended
above an infinite ground plane.
Specifically, the monopole has a length of 5 cm, radius of 1.52 mm and is suspended 1.6 mm above 
a conducting square of side length 10 cm, as shown in Fig. \ref{fig:cyl_monopole}.
To mimic an infinite ground plane, the truncating walls of the simulation domain are in direct
contact with the ends of the square.
The ground plane is coupled to the cylinder by a single, vertically oriented edge, across which is
connected a driving circuit given by a time varying voltage source connected in series to a 100$\Omega$ resistor.
The voltage fed to the resistor is assumed to be a modulated Gaussian with center frequency $f_{0}=2.5\ \text{GHz}$ and bandwidth $f_{\text{BW}}=2\ \text{GHz}$
The timestep size $\Delta_{t}$ was set to be $\left(30f_{\text{max}}\right)^{-1}$.
The mesh used to discretize the domain had an average edge length of $\left(20f_{\text{max}}\right)^{-1}$, resulting in $N_{\text{EM}}=512,436$ and the simulation was run for $N_{t}=2001$.
The setup is geometrically identical to an example in \cite{jin_riley} and looking at Fig \ref{fig:jin_monopole}, 
we see good agreement between the admittance curves generated through port extraction and a
coupled time domain solver for the same simplified probe model.
The solve time per timestep performing the extraction as detailed in Section \ref{sec:extraction} was approximately 8 seconds per timestep, with the subsequent circuit solve completing its entire run of 2001 timesteps in under $20$ ms. 

\subsection{Input impedance of a strip above a Finite Ground Plane}\label{sec:linear_benchmark}
\begin{figure}
\begin{subfigure}{0.5\textwidth}
    \centering
    \includegraphics[width=0.8\linewidth]{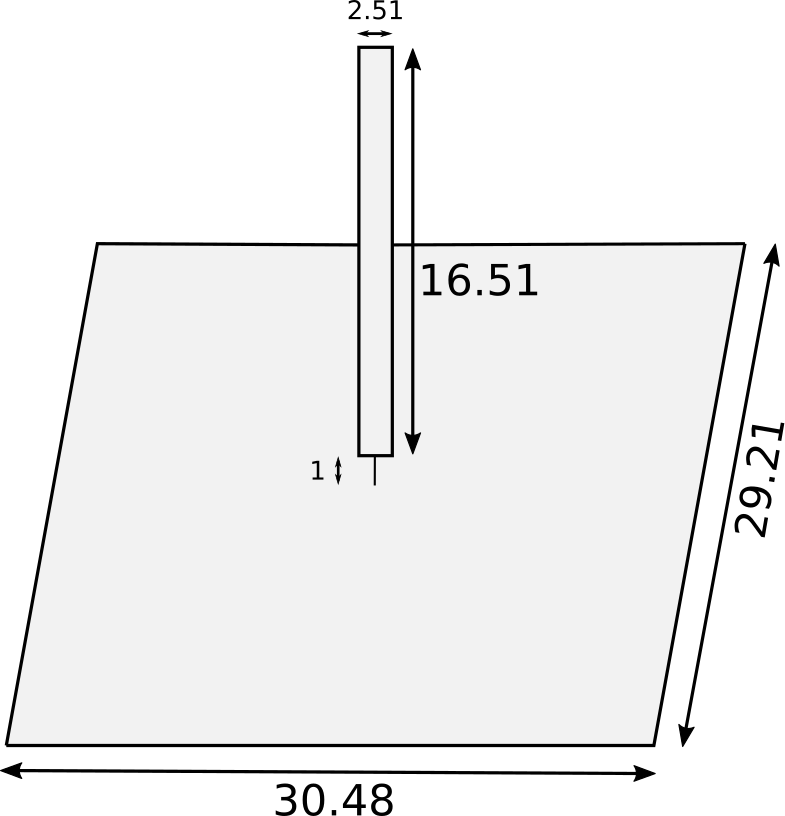}
    \caption{Geometry of the monopole antenna. The dimensions are in cm.}
    \label{fig:kong_monopole}
\end{subfigure}
\begin{subfigure}{0.5\textwidth}
    \centering
    \includegraphics[width=\linewidth]{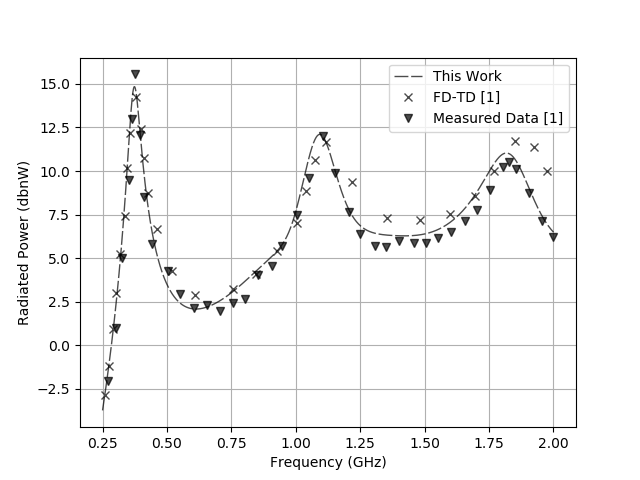}
    \caption{Power radiated due to a 1 mV source compared to measured data and FDTD \cite{kong_FDTD}}
    \label{fig:kong_power}
\end{subfigure}
\caption{Description of the geometry and obtained results for the monopole strip suspended over a finite ground plane.}
\end{figure}
We consider a conducting strip suspended over a finite ground plane, as specified in Fig. \ref{fig:kong_monopole}.
The coupling between the EM system and the driving circuit is achieved across a vertical 1 cm edge going from the conducting plane to the strip.
The circuit is assumed to be a Thevenin source characterized by $f_{0}=1$ GHz and $f_{\text{bw}}=999$ MHz connected in series to a $100\ \Omega$ resistor. 
The simulation domain is discretized using a tetrahedral mesh with approximate average edge length set to $\left(20f_{\text{max}}\right)^{-1}$, yielding $N_{\text{EM}}=2,000,936$.
The system was run for $N_{t}=4000$ timesteps (with each timestep taking approximately 13 seconds to converge) and the port parameters were extracted through Fourier transforms of the time-series data.
As is evident from Fig. \ref{fig:kong_power}, the radiated power curve shows very good agreement to measured data and FD-TD.
\subsection{Microstrip Amplifier}\label{sec:nonlinear_benchmark}
\begin{figure}
\begin{subfigure}{0.5\textwidth}
    \centering
    \includegraphics[width=\linewidth]{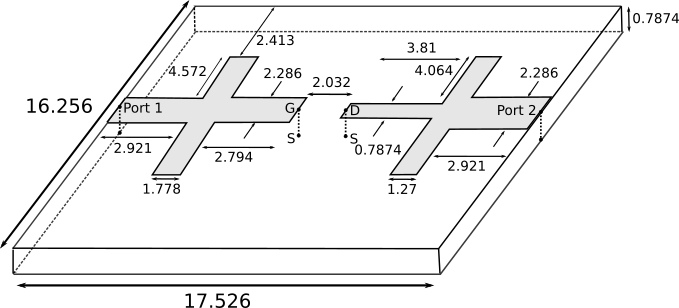}
    \caption{Geometry of the microstrip amplifier with a FET attached between G and D \cite{wang_jin_2008}. The dimensions are in mm.}
    \label{fig:microwave_scheme}
\end{subfigure}
\begin{subfigure}{0.5\textwidth}
    \centering
    \includegraphics[width=\linewidth]{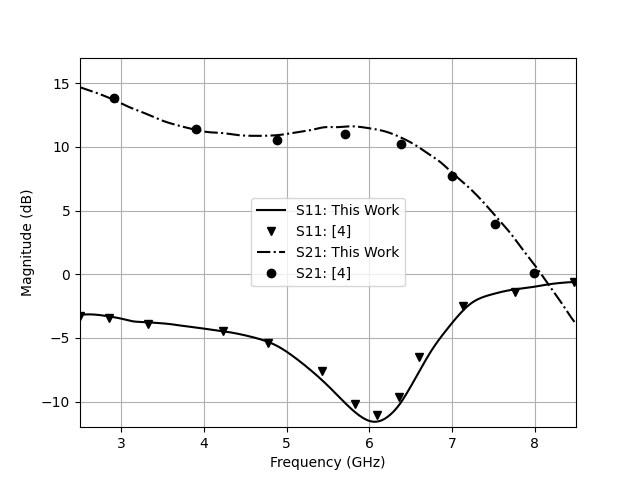}
    \caption{Comparison of $S_{11}$ and $S_{21}$ for a microwave amplifier as described in \cite{wang_jin_2008}}
    \label{fig:microwave_amp_S}
\end{subfigure}
\caption{Geometry description and calculated $S$-parameters for the microwave amplifier.}
\end{figure}
Next, we validate the proposed technique for nonlinear circuit systems by comparing the reflection coefficient and gain for a microstrip amplifier.
The geometry and driving circuits are exactly as in \cite{wang_jin_2008} and we obtain the \textit{S} parameters through small signal analysis, with $f_{0}=5.5\ \text{GHz}$, $f_{\text{bw}}=3.5\ \text{GHz}$ and $f_{\text{max}}=f_{0}+f_{\text{bw}}$.
The tetrahedral mesh used to discretize the domain has an average edge length of $\left(15f_{\text{max}}\right)^{-1}$ with $N_{\text{EM}}=5,134,732$.
The data used to compute the scattering parameters was obtained by running this setup for $N_{t}=6000$ timesteps.
We note from Fig. \ref{fig:microwave_amp_S} that the measured \textit{S} parameters show good agreement to results from \cite{wang_jin_2008}.
Extracting this response took approximately 37 seconds per timestep, and the nonlinear circuit solve completed in just over 5 seconds.

\subsection{Miscrostrip Rectifier modelled through Drift-Diffusion}
\begin{figure}
    \begin{subfigure}{0.5\textwidth}
    \centering
    \includegraphics[width=\linewidth]{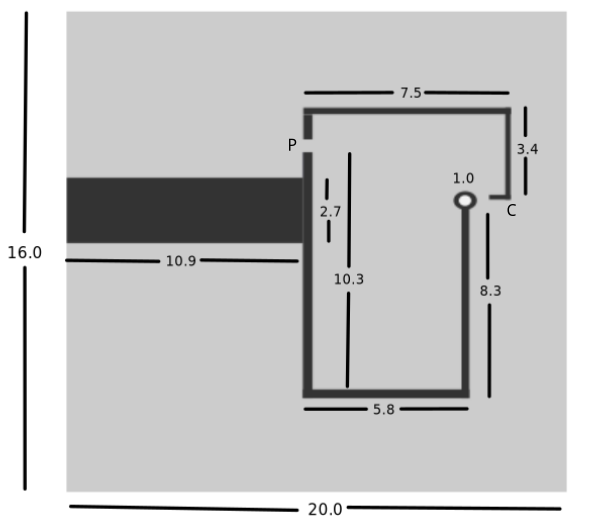}
    \caption{Schematic of the microstrip rectifier. All dimensions are in mm.}
    \label{fig:ADS_v_pm_schematic}
    \end{subfigure}
    \begin{subfigure}{0.5\textwidth}
    \includegraphics[width=\linewidth]{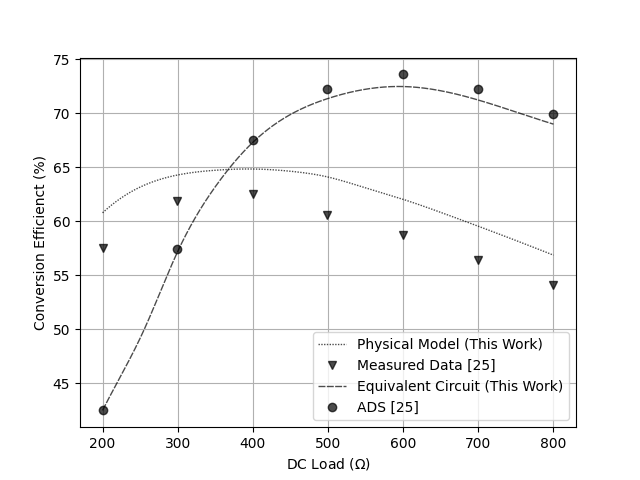}
    \caption{Computed conversion efficiency for a diode rectifier with a simulation done through a physical model compared to ADS \cite{schottkey}.}
    \label{fig:ADS_v_pm}
    \end{subfigure}
    \caption{Geometry layout and comparison of conversion efficiency for a microstrip rectifier circuit.}
\end{figure}
Until now we have demonstrated the use of port extraction on linear and nonlinear systems connected to nodal circuit networks. Next, we aim to show that the proposed method works with systems where the devices are governed nonlinear differential equations. The system under analysis is a microstrip rectifier circuit as shown in Fig. \ref{fig:ADS_v_pm_schematic}. The thickness of the board was 1 mm and the relative permittivity of the substrate 2.65.
A HSMS-282B diode is placed across port $P$ (with physical parameters for \eqref{eq:DDESchottky} as in \cite{schottkey}) with a $10$ pF filter capacitor attached to a variable load across port $C$.
The input source was assumed to be a modulated Gaussian with $f_{0}=2.45\ \text{GHz}$ and $f_{\text{bw}}=0.25\ \text{GHz}$.
We performed two experiments on a microstrip rectifier circuit: (1) First, the Schottky diode in the layout was modeled using an equivalent circuit network, mimicking a similar setup simulated on ADS. (2) Next, \emph{using the same extracted port response as in the first experiment}, we modeled the diode using a set of Drift-Diffusion \cite{schottkey} equations in \eqref{eq:DDESchottky}.
In each case, the conversion efficiency of the rectifier 
\begin{equation}
    \eta = \frac{P_{\text{DC}}}{P_{\text{source}}}\cdot 100 \%
\end{equation}
where $P_{\text{DC}}$ denotes the power measured at the output end $P_{\text{source}}$ the corresponding quantity at the soruce was compared against data from \cite{schottkey}. As is evident in Fig. \ref{fig:ADS_v_pm}, in the first experiment, results obtained through the proposed method agree well with corresponding results obtained through ADS EM Co-simulation. In the second experiment, our results better match measured data of the rectifier circuit than the corresponding efficiency curve predicted by ADS. We emphasize the fact that the results from the equivalent circuit \emph{do not agree} with experimental measurements, due to the network not being representative of the actual diode for the parameters chosen, thereby illustrating a situation where the ability to couple the EM layout with a general device model is a significant advantage.
\subsection{Strip above a Finite Ground Plane driven by different circuits}\label{sec:showing_off}

In keeping with objectives stated earlier, we first extracted the port parameters following the procedure in
Section \ref{sec:extraction} for the example used in Section \ref{sec:linear_benchmark} with $\Delta_{t}=16\ \text{ps}$. 
This extracted response was then attached to a Chebyshev filter and a Diode Mixer circuit 
respectively, and the obtained port voltages in time were compared to equivalent results obtained from a direct solution of the coupled system.
\subsubsection{Chebyshev filter}
\begin{figure}
\begin{subfigure}{.5\textwidth}
    \centering
    \begin{circuitikz}[scale=0.5,transform shape]
       \draw (0,0) to [sV, l=\Large $V_s(t)$,*-*] (0,3);
       \draw (3,0) to [capacitor, l=\Large$ 9.05$ pF, *-*]      (3,3);
       \draw (6,0) to [capacitor, l=\Large$13.48$ pF, *-*]      (6,3);
       \draw (9,0) to [capacitor, l=\Large$ 9.05$ pF, *-*]      (9,3);
       \draw (0,0) to [short]          (6,0);
       \draw (6,0) to [short,*-*]          (12,0);
       \draw (9,3) to [short,*-*]          (12,3);
       \draw (0,3) to [R, l=\Large$50$ $\Omega$, *-]       (3,3);
       \draw (3,3) to [inductor, l=\Large$16.308$ nH, *-]       (6,3);
       \draw (6,3) to [inductor, l=\Large$16.308$ nH, *-]       (9,3);
       \draw (12,0) to [open, l=\Large$V_{\text{port}}(t)$] (12,3);
       \draw (12,0) to [open, l=\Large $+$] (12,5);
       \draw (12,0) to [open, l=\Large $-$] (12,1);
    \end{circuitikz}
    \caption{Schematic of the Chebyshev filter used.}
    \label{fig:kong_chebyshev}
\end{subfigure}
\begin{subfigure}{.5\textwidth}
    \centering
    \includegraphics[width=\linewidth]{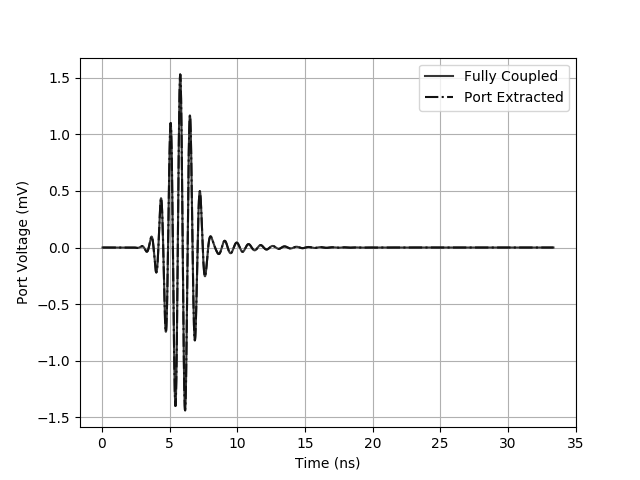}
    \caption{Plot of the port voltage for the Chebyshev filter from Fig. \ref{fig:kong_chebyshev} obtained through the fully coupled and extracted responses.}
    \label{fig:kong_lin_result}
\end{subfigure}
\caption{Circuit description and comparison of port voltages between the port extraction and fully coupled methods for a linear circuit system.}
\end{figure}
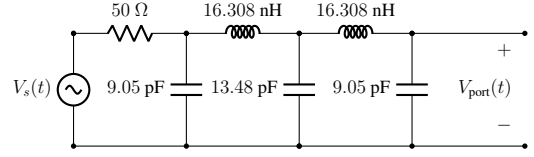
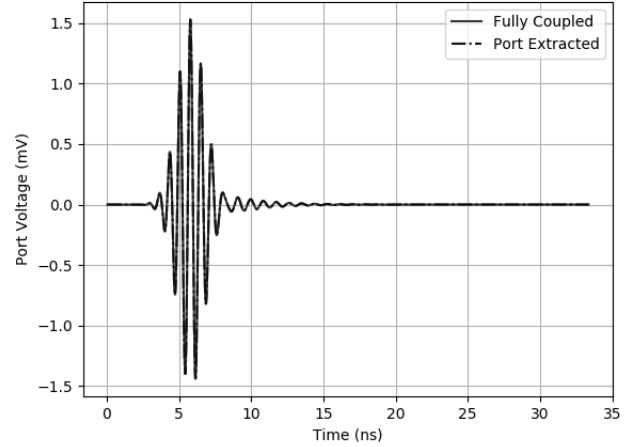
First, we use the \emph{extracted} transient port parameters on a Chebyshev filter as shown in Fig. \ref{fig:kong_chebyshev}. $V_{s}(t)$ was characterized by $f_0=1.5\ \text{GHz}$ and $f_{\text{bw}}=0.5\ \text{GHz}$. 
The timestep size in the circuit system was set to the same size used in the extraction of the EM response.
The comparison of the port voltages as a function of time are shown in Fig. 
\ref{fig:kong_lin_result}.
The $L^{2}$ error between the two solutions was $3.1\times 10^{-12}$.

\subsubsection{Diode Mixer}\label{sec:mixer}
\begin{figure}
\begin{subfigure}{.5\textwidth}
    \centering
    \begin{circuitikz}[scale=0.3,transform shape, american voltages]
  		\draw (0,0) to [sV,l=\Large$\text{RF}$,*-*] (0,3)node[label={[font=\footnotesize]above:\Large$1$}] {};
  		\draw (0,3) to [short]  (3,3);
  		\draw (3,3) to [R, l=\Large$1$ M $\Omega$ , *-] (3,0);
  		\draw (3,3) to [R, l=\Large$100$ $\Omega$, *-*] (6,3)node[label={[font=\footnotesize]above:\Large$2$}] {};
  		\draw (6,3) to [R, l=\Large$100$ $\Omega$, *-] (6,0);
  		\draw (6,3) to [capacitor, l=\Large$1$ pF, *-*] (9,3)node[label={[font=\footnotesize,xshift=0.2cm, yshift=0.1cm]above:\Large$3$}] {};
  		\draw (9,3) to [inductor, l=\Large$6$ nH, *-*] (9,0) node[label={[font=\footnotesize,xshift=0.2cm]above:\Large$6$}] {};
  		\draw (9,3) to [R, l=\Large$10$ $\Omega$, *-*] (9,6)    node[label={[font=\footnotesize]above:\Large$4$}] {};
  		\draw (9,0) to [R, l=\Large$10$ $\Omega$, *-*] (9,-3)   node[label={[font=\footnotesize,xshift=0.2cm]above:\Large$9$}] {};
  		\draw (9,6) to [capacitor, l=\Large $1$ pF, *-*] (12,6);
  		\draw (12,6)node[label={[font=\footnotesize]above:\Large$5$}] {} to [sV,l=\Large$\text{LO}$,*-*] (15,6);
  		\draw (9,3) to [diode,*-*] (12,3)node[label={[font=\footnotesize]above:\Large$7$}] {};
  		\draw (12,3) to [inductor,l=\Large$3$ nH,*-] (12,0);
  		\draw (12,3) to [capacitor,l=\Large$.1$  pF,*-*] (15,3);
  		\draw (15,3)node[label={[font=\footnotesize]above:\Large$8$}] {} to [resistor,l=\Large$1k\Omega$,*-*] (15,0);
  		\draw (9,-3) to [V,l=$V_{bias}$,*-] (9,-5) {};
  		\draw (0,.5) -- (0,0) node[ground]{}; 
  		\draw (3,.5) -- (3,0) node[ground]{}; 
  		\draw (6,.5) -- (6,0) node[ground]{}; 
  		\draw (12,1) -- (12,0) node[ground]{}; 
  		\draw (9,-5) -- (9,-5.5) node[ground]{};
  		\draw (15,6) -- (15,5.5) node[ground]{};  
  		\draw (15,.5) -- (15,0) node[ground]{};
		\draw (15,3) to [short,*-*]  (16,3);
		\draw (15,0) to [short,*-*]  (16,0);
	\end{circuitikz}%
    \caption{Schematic of the Diode Mixer. The EM system is attached between nodes 8 and ground}
    \label{fig:kong_mixer}
\end{subfigure}
\begin{subfigure}{.5\textwidth}
    \centering
    \includegraphics[width=\linewidth]{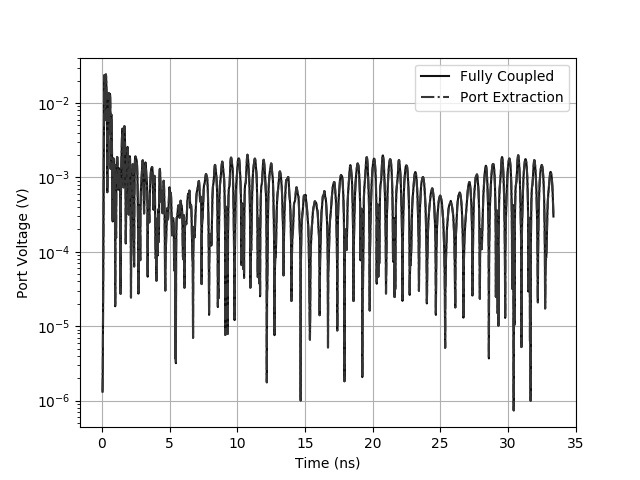}
    \caption{Plot of the port voltage for the Diode Mixer from Fig. \ref{fig:kong_mixer} obtained through the fully coupled and extracted responses.}
    \label{fig:kong_nonlin_result}
\end{subfigure}
\caption{Circuit description and comparison of port voltages between the port extraction and fully coupled methods for a nonlinear circuit system.}
\end{figure}
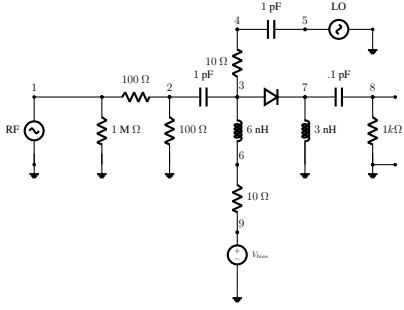
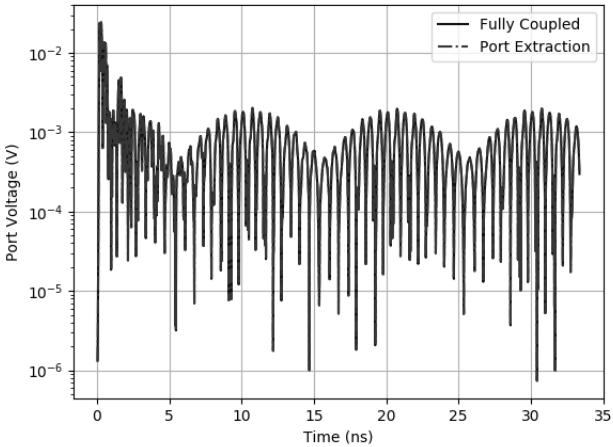
Next, we use extracted port parameter with a nonlinear Diode Mixer as shown in Fig. \ref{fig:kong_mixer}.
The diode between nodes 3 and 7 has a saturation current $I_{s}=2\ \text{nA}$, emission 
coefficient $\eta=2.0$ and $k_{B}T/q=25.6\ \text{mV}$. 
The RF and LO sources were assumed to be sine waves of magnitude 0.4 V with frequencies 900 MHz and 800 MHz respectively.
The current across the diode was modelled using the Shockley 
equation and the bias voltage was set to 0.7 V to activate the diode.
The relative $L^{2}$ error between the two curves in Fig. \ref{fig:kong_nonlin_result} was $4.7\times 10^{-12}$.

\subsubsection{Computational Complexity}
\begin{figure}
    \centering
    \includegraphics[width=\linewidth]{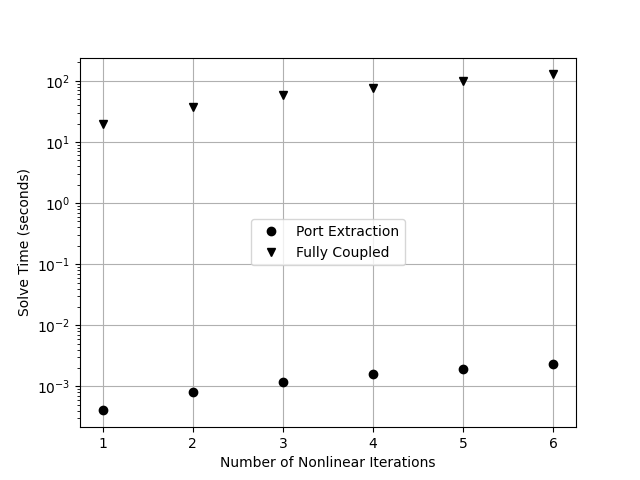}
    \caption{Cumulative solution time of the linearized system from Section \ref{sec:mixer} as a function of the number of iterations.}
    \label{fig:kong_complexity}
\end{figure}

Finally, we discuss the asymptotic cost complexity of the proposed method. Let $N_{\text{EM}}$ and $N_{\text{CKT}}$ denote the number of degrees of freedom of the EM and circuit systems, respectively;
$N_{t}$ the number of timesteps; $N_{\text{NL}}$ the number of nonlinear iterations per time step; $N_{p}$ the number of circuit ports and $N_{\text{GMRES}}$ the number of matrix multiplications required to solve the linearized system denoted in the superscript.
The cost for solving the fully coupled system is
\begin{equation}\label{eq:cost_coupled}
    C_{\text{coupled}} = \mathcal{O}\left (N_{t}N^{\text{coupled}}_{\text{NL}}N^{\text{coupled}}_{\text{GMRES}}\left(N_{\text{EM}}+N_{\text{CKT}}\right) \right).
\end{equation}
On the other hand, the cost of port extraction is
\begin{equation}\label{eq:cost_pe}
    \begin{split}
    C_{\text{PE}} &= C_{\text{PE,1}} + C_{\text{PE,2}} \\
    &= \mathcal{O}(N_{t}N_p N^{\text{PE}}_{\text{GMRES}}N_{\text{EM}})
    +\mathcal{O}(N_{t}N^{\text{PE}}_{\text{NL}}N^{\text{CKT}}_{\text{GMRES}}N_{\text{CKT}}).
    \end{split}
\end{equation}
Before, we proceed, we note the following. Typically, $N_{\text{EM}} \gg N_{\text{CKT}}$. As a result, we ignore the cost of computing the Jacobian in \eqref{eq:cost_coupled}. In \eqref{eq:cost_pe}, the first portion refers to the cost of exciting each port and obtaining the corresponding response at other ports. It is a one-time cost and not incurred as one marches through (indeed, it is the characteristic of the EM systems and circuit agnostic). The second term in \eqref{eq:cost_pe}, is the cost of non-linear solve at each port.  Typically, the number of non-linear solves, $N_{\text{NL}}^{\text{coupled}} N^{\text{coupled}}_{\text{GMRES}}$ in \eqref{eq:cost_coupled}, is significantly larger than $N_{p}N^{\text{PE}}_{\text{GMRES}}$ as it involves a fully coupled solve involving all the degrees of freedom in the system. 
In order to meaningfully compare computational costs, it is important to incorporate the contributions of both $C_{\text{PE,1}}$ and $C_{\text{PE,2}}$ against $C_{\text{coupled}}$.
To do this, we considered the solve time per nonlinear iteration within a single timestep of each solve for the example in Section \ref{sec:mixer}, i.e the finite ground plane monopole antenna driven by a diode mixer circuit. This includes the cost for evaluating a single timestep in the extraction process as well as the cost of computing $N_{\text{NL}}$ nonlinear iterations, where $N_{\text{NL}}$ is the average number of nonlinear iterations required to achieve convergence per timestep. In this example,  $N_{\text{NL}}$= 6, per time step. Extracting the impulse response took 13s per $N_t$, i.e., $C_{\text{PE,1}} = 13N_t$. What is compared in  Fig. \ref{fig:kong_complexity} are $C_{\text{PE,2}}/N_t$ and $C_{\text{Coupled}}/N_t$. As is evident, $C_{\text{PE,2}} \ll C_{\text{coupled}}$.



\section{Conclusion}

In this paper, we have demonstrated a technique to extract transient port parameters from a coupled EM-circuit solver. We have shown  that the technique is stable, circuit agnostic, computationally efficient and produces solutions that are \emph{numerically identical} to those obtained through a traditional fully coupled solve.
The extension of this method to more sophisticated domain-decomposition solvers, application to MIMICs and resolution of continuous nonlinear material distributions in the EM system will be explored in subsequent papers.

\section*{Acknowledgments}

This work was supported by SMART Scholarship program, the Department of Energy Computational Science Graduate Fellowship under grant DE-FG02-97ER25308 and from the NSF via CMMI-1725278. Balasubramaniam would like to acknowledge conversations with Prof. Jianming Jin that inspired this work and \cite{scott_pe}. 
\bibliographystyle{IEEEtran}
\bibliography{bibliography}

\begin{IEEEbiography}[{\includegraphics[width=1in,height=1.25in,clip,keepaspectratio]{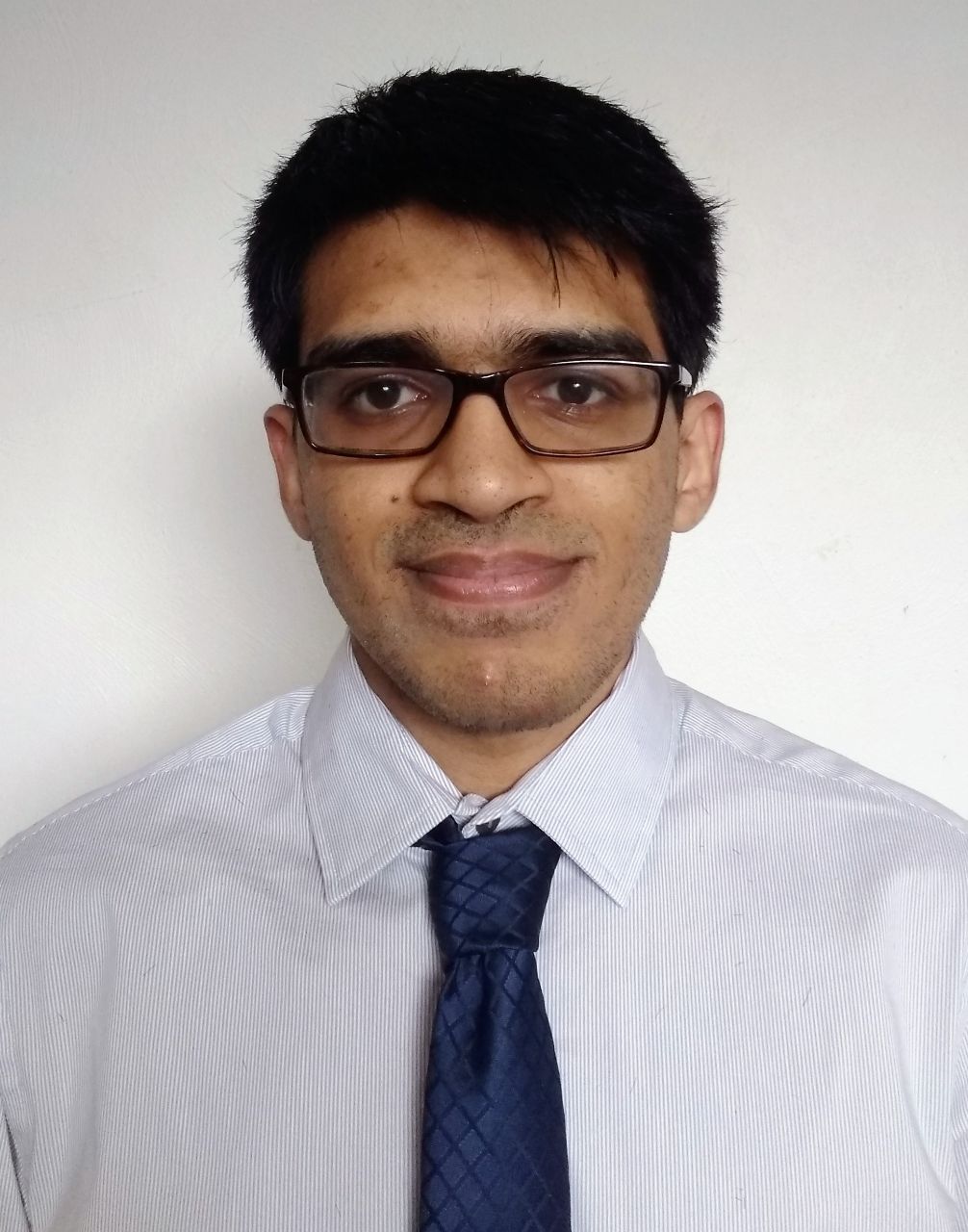}}]{Omkar H. Ramachandran} (S'20) received his B.A. degree
in physics from the University of Colorado at Boulder in 2018 and is currently pursuing his Ph.D in electrical and computer engineering at Michigan State University, East Lansing, MI. His research interests include several topics in computational electromagnetics, including particle-in-cell methods, analysis of coupled EM-device systems and nonlinear optimization.
\end{IEEEbiography}
\begin{IEEEbiography}[{\includegraphics[width=1in,height=1.25in,clip,keepaspectratio]{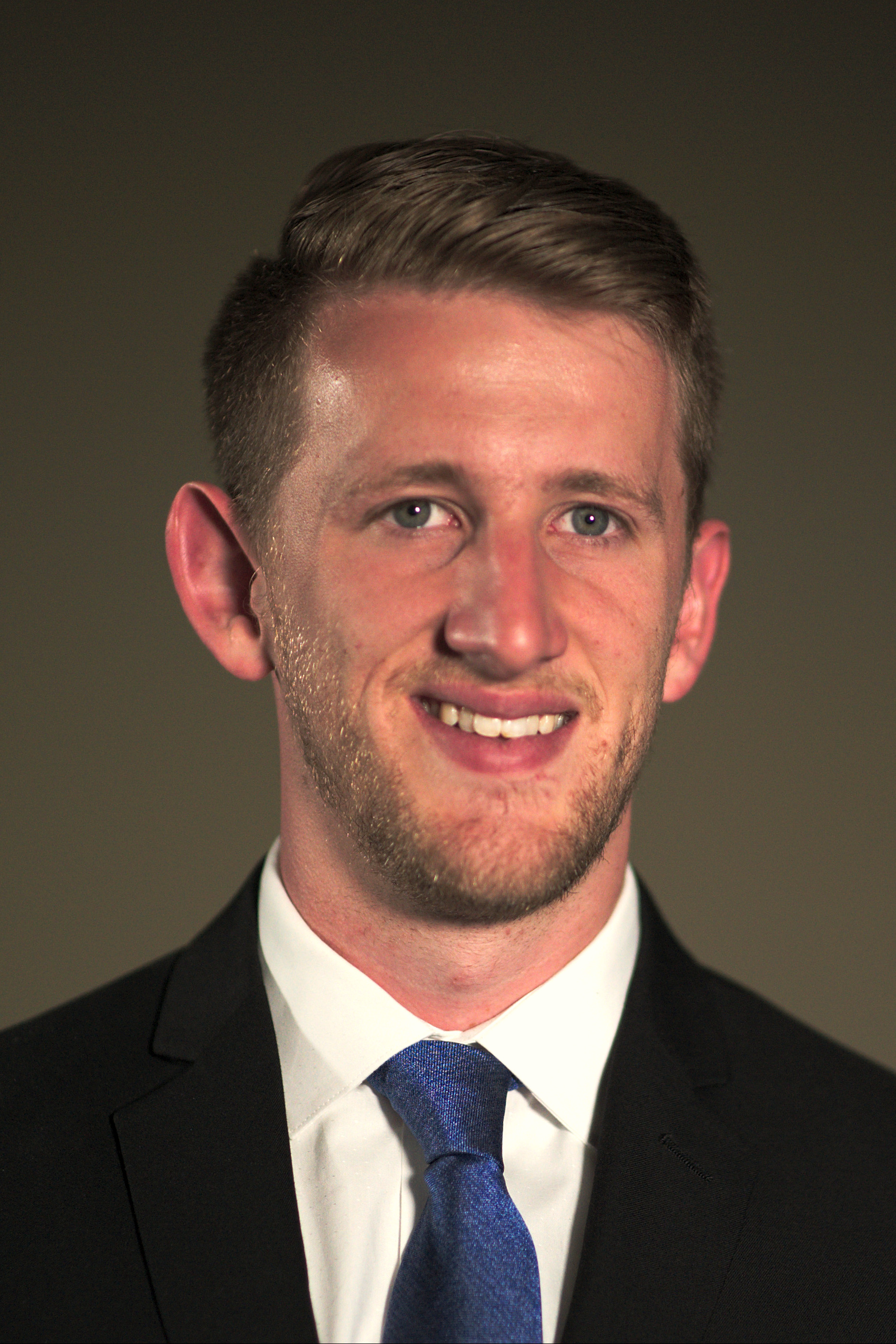}}]{Scott O'Connor} (S'13) received the B.S. degrees
in electrical engineering, M.S. degree in electrical
engineering and Ph.D. degree in computational elec-
tromagnetics from Michigan State University, East
Lansing, MI, USA, in 2014, 2017 and 2021 respec-
tively. His research interests include coupled circuit
electromagnetic solvers, finite element methods and
particle in cell methods.
\end{IEEEbiography}

\begin{IEEEbiography}[{\includegraphics[width=1in,height=1.25in,clip,keepaspectratio]{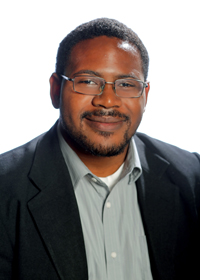}}]{Zane D. Crawford} (S'15) received the B.S. degree
in electrical engineering and computer engineering
from Michigan State University, East Lansing, MI,
USA. He is currently pursuing the Ph.D. degree in
computational electromagnetics.
His research interests include several aspects
of computational electromagnetics, including time
and frequency domain finite element methods and
particle-in-cell methods.
Mr. Crawford was a recipient of the Department of
Energy Computational Science Graduate Fellowship in 2015.
\end{IEEEbiography}
\begin{IEEEbiography}[{\includegraphics[width=1in,height=1.25in,clip,keepaspectratio]{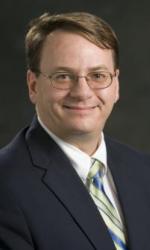}}]{Leo Kempel}
(S’89–M’94–SM’99–F’09) was
born in Akron, OH, USA, in 1965. He received the
B.S.E.E. degree from the University of Cincinnati,
Cincinnati, OH, USA, in 1989, and the M.S.E.E.
and Ph.D. degrees from the University of Michigan, Ann Arbor, MI, USA, in 1990 and 1994,
respectively.
After a brief post-doctoral appointment at the
University of Michigan, he joined Mission Research
Corporation, Goleta, CA, USA, in 1994, as a Senior
Research Engineer. He led several projects involving
the design of conformal antennas, computational electromagnetics, scattering
analysis, and high-power/ultrawideband microwaves. He joined Michigan
State University, East Lansing, MI, USA, in 1998. He served as an IPA
with the Air Force Research Laboratory’s Sensors Directorate, Riverside,
OH, USA, from 2004 to 2005 and 2006 to 2008. He was the Inaugural
Director of the Michigan State University High Performance Computing
Center, East Lansing. He was the first Associate Dean for Special Initiatives
with the College of Engineering, Michigan State University, from 2006 to
2008, and the Associate Dean for Research from 2008 to 2013. He then
became the Acting Dean of Engineering in 2013. Since 2014, he has been
the Dean with the College of Engineering, Michigan State University. He has
co-authored the book The Finite Element Method for Electromagnetics (IEEE
Press). His current research interests include computational electromagnetics,
conformal antennas, microwave/millimeter-wave materials, and measurement
techniques.
Prof. Kempel is a fellow of the Applied Computational Electromagnetics
Society (ACES). He was a member of the Antennas and Propagation Society’s
Administrative Committee and the ACES Board of Directors. He is a member
of Tau Beta Pi, Eta Kappa Nu, and Commission B of URSI. He served as
the Technical Chairperson for the 2001 ACES Conference and the Technical
Co-Chair for the Finite Element Workshop held in Chios, Greece, in 2002.
He was the Fellow Evaluation Committee Chairperson for the IEEE Antennas
and Propagation Society and served on the IEEE Fellow Board from 2013 to
2015. He was a recipient of the CAREER Award by the National Science
Foundation, the Teacher-Scholar Award by Michigan State University in
2002, and the MSU College of Engineering’s Withrow Distinguished Scholar
(Junior Faculty) Award in 2001. He served on the U.S. Air Force Scientific
Advisory Board from 2011 to 2015. He served as an Associate Editor of
the IEEE TRANSACTIONS ON ANTENNAS AND PROPAGATION. He is an
active reviewer for several IEEE publications as well as the Journal of
Electromagnetic Waves and Applications and Radio Science.
\end{IEEEbiography}
\begin{IEEEbiography}[{\includegraphics[width=1in,height=1.25in,clip,keepaspectratio]{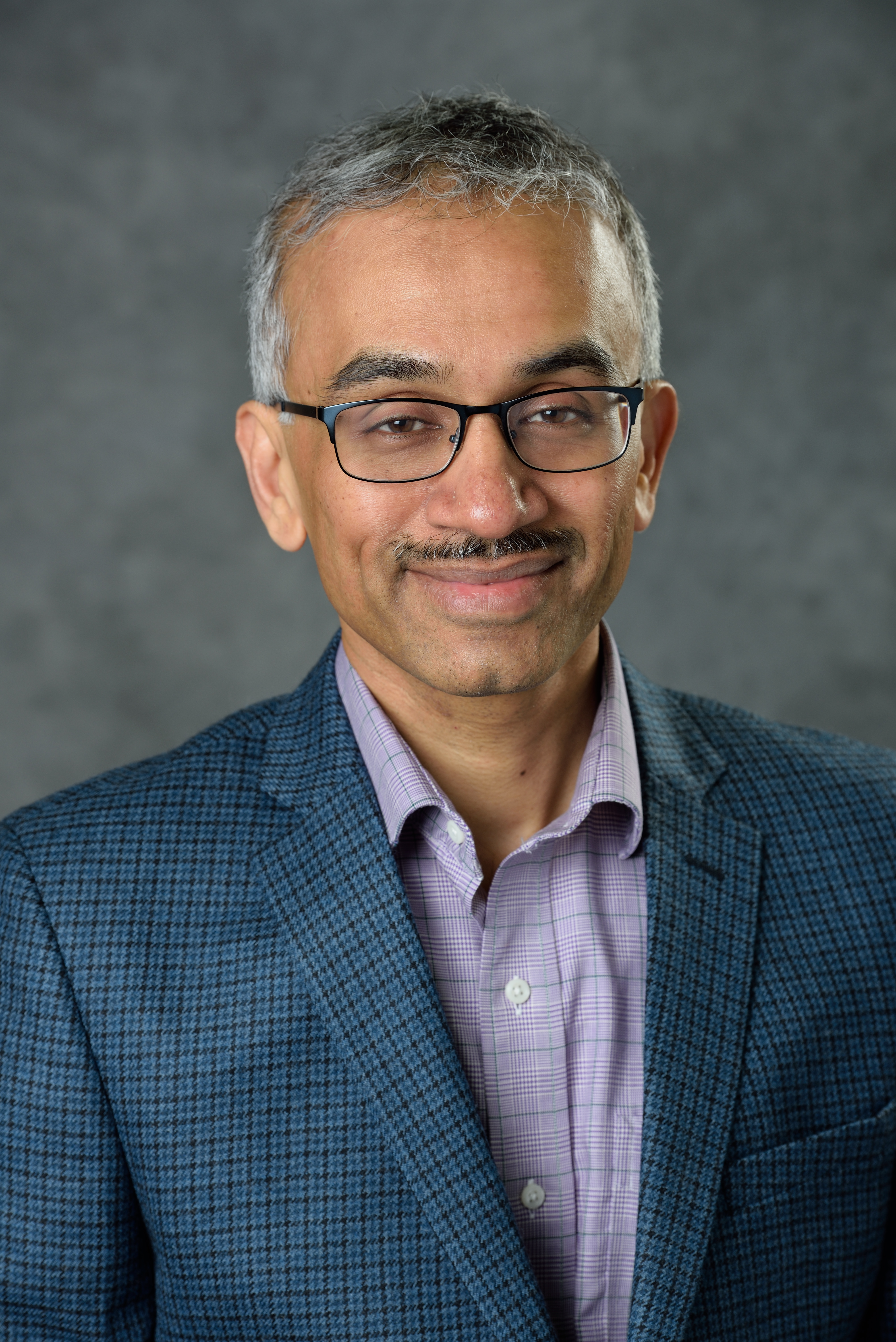}}]{B. Shanker} received his B'Tech from the Indian Institute of Technology, Madras, India in 1989, M.S. and Ph.D in 1992 and 1993, respectively, from the Pennsylvania State University. From 1993 to 1996 he was a research associate in the Department of Biochemistry and Biophysics at Iowa State University where he worked on the Molecular Theory of Optical Activity. From 1996 to 1999 he was with the Center for Computational Electromagnetics at the University of Illinois at Urbana-Champaign as a Visiting Assistant Professor, and from 1999-2002 with the Department of Electrical and Computer Engineering at Iowa State University as an Assistant Professor. Currently, he is a University Distinguished Professor (an honor accorded to about 2\% of MSU faculty members) in the Department of Electrical and Computer Engineering at Michigan State University, and the Department of Physics and Astronomy. From 2015-2018,  he was appointed Associate Chair of the Department of Computational Mathematics, Science and Engineering, a new department at MSU and was a key player in building this Department.  Earlier he served as the Associate Chair for Graduate Studies in the Department of Electrical and Computer Engineering from 2012-2015, and currently is the Associate Chair for Research in ECE. He has authored/co-authored around 450 journal and conference papers and presented a number of invited talks. His research interests include all aspects of computational electromagnetics (frequency and time domain integral equation based methods, multi-scale fast multipole methods, fast transient methods, higher order finite element and integral equation methods), propagation in complex media, mesoscale electromagnetics, and particle and molecular dynamics as applied to multiphysics and multiscale problems. He was an Associate Editor for IEEE Antennas and Wireless Propagation Letters (AWPL) and IEEE Transactions on Antennas and Propagation, was the Topical Editor for Journal of Optical Society of America: A, and is a full member of the USNC-URSI Commission B. He is Fellow of IEEE, elected for his contributions in computational electromagnetics. He has also been awarded the Withrow Distinguished Junior scholar (in 2003), Withrow Distinguished Senior scholar (in 2010), the Withrow teaching award (in 2007), and the Beal Outstanding Faculty award (2014)
\end{IEEEbiography}

\end{document}